\documentclass[12pt]{article}

\usepackage{amsmath}
\usepackage{amssymb}

\begin{document}

\title{Relation between the critical spin and angular velocity of a nucleus
immediately after backbending
\footnote{Zh. Eksp. Teor. Fiz. {\bf 76}, 1506--1514 (1979) [Sov. Phys. JETP
{\bf 49,} No.~5, 765--769 (1979)]}}

\author{V.G. Nosov$^{\dagger}$ and A.M. Kamchatnov$^{\ddagger}$\\
$^{\dagger}${\small\it Russian Research Center Kurchatov Institute, pl. Kurchatova 1,
Moscow, 123182 Russia}\\
$^{\ddagger}${\small\it Institute of Spectroscopy, Russian Academy of Sciences,
Troitsk, Moscow Region, 142190 Russia}
}

\maketitle

\begin{abstract}
In nonspherical nuclei at $J = J_c + 0$ the relationship between the angular
momentum and angular velocity immediately after backbending is the same as in
the limiting case $J - J_c\to\infty$. This indicates that there is a unique type
of cancellation of the deviations from a rigid-body moment of inertia in the
upper phase $J>J_c$. An integral relationship is found which expresses this
cancellation quantitatively. This formula permits $J_c$ to be calculated for the
rotational bands of the even-even nuclei studied and the results are in agreement
with those obtained by other methods of locating the Curie point. For the ground
state band of W$^{170}$ the cancellation of the reciprocals of the true and
rigid-body moments of inertia can be verified directly. The condition for the
stability of the rotation of a nonspherical nucleus is analyzed in the Appendix
in close connection with the problem of a reasonable definition of the concept
of a variable moment of inertia.
\end{abstract}

\section{Introduction}

Even before the discovery in 1971 of the singularity exhibited by the rotational
band at a certain critical value of the nuclear spin $J = J_c$, it was gradually
becoming clear that we are actually dealing with a situation where the moment of
inertia $I$ of a nonspherical nucleus varies most significantly.   In choosing a
reasonable definition of this concept it is desirable to keep in mind the
following considerations, in addition to purely aesthetic ones. First of all,
for completeness and internal consistency of the theory it is important that both
definitions of the moment of inertia, that via the Lagrangian and that via the
Hamiltonian, be equivalent.   In addition, in accordance with general physical
considerations it is natural to expect that precisely at negative values of the
correctly defined moment of inertia the rotation becomes unstable; this is
discussed in more detail in the Appendix.

The following definition meets all these requirements:
\begin{equation}\label{1}
    \hbar\Omega=\frac{dE}{dJ},\qquad \frac{\hbar^2}I=\frac{d(\hbar\Omega)}{dJ}=
    \frac{d^2E}{dJ^2}
\end{equation}
(the notation is the same as in Ref.~1).   We emphasize that these prerequisites
do not create any practical difficulties in a concrete comparison with experiment
since in formulas like (1) it is possible to replace the derivatives with respect
to $J$ by the corresponding directly observable finite differences.   In fact,
the widely known Bohr-Mottelson formula
\begin{equation}\label{2}
    E\cong\frac{\hbar^2}{2I'}J(J+1)
\end{equation}
has the property that the finite differences calculated with it agree with the
result of the formal differentiation according to (1).   On the other hand, for
$J\gg 1$ the replacement of the derivatives by finite differences suggests itself
automatically and no special problems appear.

In our earlier study [1] this phenomenon was viewed as a smooth, continuous
rearrangement of the angular momentum coupling scheme in the nucleus. For adiabatically
slow rotation, $J\ll J_c$, the internal state of the system is formed mainly by a
``nucleon-nuclear symmetry axis" type of interaction, which is due to the axially
symmetric deformation of a nucleus ``at rest".   However, in reality the nucleus is
rotating, so that the nucleon-rotation axis interaction (which here emerges as a
certain non-adiabatic correction) is always effective to some degree.   In the
entire region $J <J_c$ there is a complicated ``nucleon-nuclear symmetry axis" plus
``nucleon-rotation axis" coupling scheme of relatively low symmetry and the two
interactions, generally speaking, are comparable.   However, in a sufficiently strong
rotation ``field" the mechanical angular momenta of the individual quasiparticles are
aligned parallel to the $\mathbf{\Omega}\|\mathbf{J}$ direction and cease to be
oriented along the vector $\mathbf{n}$.   It can be said that this corresponds to
the simplest, most symmetric nucleon-rotation axis coupling scheme, not directly
affecting the direction $\mathbf{n}$ of the nuclear axis. Then the vector n remains
``free," that is, it is actually distributed isotropically for $J\geq J_c$.
As a consequence of the increased symmetry of the rotational state at the point
$J =J_c$ the angular velocity of the rotation falls abruptly by some amount
$\Delta(\hbar\Omega)$ and in the isotropic upper phase the moment of inertia
displays the seemingly paradoxical limiting behavior
\begin{equation}\label{3}
    I\cong j/(J-J_c),\qquad J-J_c\ll j/I_0
\end{equation}
($j$ is some coefficient depending on the particular nucleus and $I_0$ is the rigid
body moment of inertia).

The theory that we developed earlier [1] does not claim to be able to calculate the
specific values of such parameters as $J_c$, the discontinuity $\Delta(\hbar\Omega)$,
or the coefficient $j$ for individual nuclei.   In relation to this it should also
be noted that the less symmetric lower phase is considerably more difficult to study
theoretically.   The main result here is the square-root law
\begin{equation}\label{4}
    Q\propto (J_c-J)^{1/2},
\end{equation}
according to which the static quadrupole moment vanishes near the Curie point.
It seems almost certain that it is not possible to quantitatively determine the
value of the critical angular velocity $\Omega_{mc}$, for example, from only
deductive considerations.   However, it is remarkable that for a given $J_c$
the value of $\Omega_{nc}$ to which the rotational velocity falls after its
discontinuous decrease, can be calculated in a closed form.

\section{The angular momentum and angular velocity in the upper phase $J>J_c$}

After reaching the rigid body value of the moment of inertia
\begin{equation}\label{5}
    I=I_0, \qquad J-J_c\gg j/I_0,
\end{equation}
the rotational motion of the two components of nuclear matter can be assumed to be
fully concurrent.   Let us analyze this rotation in detail, appealing to a physical
manifestation of the proton component like the magnetic moment.   It is obvious that
here
\begin{equation}\label{6}
    g=g_0=\frac{Z}A
\end{equation}
is valid for the gyromagnetic factor.\footnote{It is well known thai in adiabatic
region the nuclear matter is taken into rotation only partially and the proton
component is involved into rotation less than the neutron component.For the
lowest levels of even-even nuclei it is observed experimentally that $g<g_0$;
see, e.g.~[2]. This physical non-equivalence of protons and neutrons must
disappear in the solid state limit of rotation. It is not necessary to
take into account spin magnetism of separate nucleons in our macroscopic approach.}

However, the nuclear magnetic moment can be calculated in a somewhat different way,
using the Larmor theorem (see Ref.~[3], for example) in, so to speak, its
differential form.   If it is favorable for a spherical nucleus to rotate in the
above manner, then in an applied magnetic field $H$ it can be viewed as rotating,
but with an angular velocity decreased by the Larmor value
\begin{equation}\label{7}
    \Omega_L=\frac{g_0eH}{2m_pc}
\end{equation}
($m_p$ is the proton mass).   Therefore, the rotational properties of the system that
we are interested in here are described by the function $E(\Omega)$, while the change
of energy is $-\Omega_LdE/d\Omega$.   According to the usual point of view, the
nucleus at a level $J$ is a particle having magnetic moment $\mu$ and additional
energy $-\mu H$ in the field (both the angular momentum vector and the magnetic
field vector are assumed to be directed along the $z$ axis).

Let us equate the two expressions for the additional energy of the system in the
magnetic field:
\begin{equation}\label{8}
-\frac{dE}{d\Omega}\Omega_L=-\mu H.
\end{equation}
Furthermore, transforming the derivatives according to (1) and substituting formula (7),
we also take into account the fact that according to the conventional definition of
the nuclear $g$ factor, the nuclear magneton $e\hbar/2m_pc$ must serve as the unit
of measurement of the magnetic moment:
\begin{equation}\label{9}
    g=g_0I\Omega/\hbar J.
\end{equation}
Finally, substituting here the values (5) and (6), we find
\begin{equation}\label{10}
    \hbar J=I_0\Omega,\qquad J-J_c\gg j/I_0.
\end{equation}

This seemingly natural recovery of the simple proportionality between the angular
momentum and the angular velocity sheds additional light on the physical nature of
the upper phase.   For $J -J_c\gg j/I_0$ its characteristics cannot depend on such
non-universal parameters as the coefficient $j$, for example.  Roughly speaking,
here we have a fairly clear interpretation of the ordinary rotation of a rigid body.
However, if we move down the band in the negative direction of the $J$ axis the
stability of this regime deteriorates.   In close relation to this the moment of
inertia undergoes an odd, non-monotonic change. First it falls off, then passes
through a minimum, and then approaches the pole according to (3).   In the region
near the transition point $J-J_c\lesssim j/I_0$ of the variable moment of inertia
the rotation, of course, can no longer be viewed as simply rigid body rotation.

Let us return for a while to the region where the moment of inertia is constant
$I=I_0$. In our earlier study [1] in a somewhat formal manner we found
\begin{equation}\label{11}
    \hbar\Omega=\hbar\Omega_{nc}+\hbar^2(J-J_c)/I_0
\end{equation}
for the rotational velocity. Now, comparing this to formula (10), we have
\begin{equation}\label{12}
    \hbar J_c=I_0\Omega_{nc}.
\end{equation}

Therefore, in reaching the pole $J = J_c + 0$ the effect of the above-mentioned
non-monotonic change of the moment of inertia on the rotational velocity is
cancelled.  It is easy to find an integral relation expressing this
cancellation in a quantitative form.   For this according to the second formula
in (1) we integrate the inverse value of the moment of inertia from $J_c$ up to
some large $J$; then, performing an even more trivial integration of the analogous
expression with the rigid body moment of inertia $I_0$, we take into account
relation (12) and the fact that for $J\to\infty$ equation (10) is valid.  As a
result we easily find
\begin{equation}\label{13}
    \int_{J_c}^\infty\left(\frac1{I_0}-\frac1I\right)dJ=0.
\end{equation}
For practical purposes it is convenient to write this relation in a dimensionless
form:
\begin{equation}\label{14}
\int_{J_c}^\infty(1-I_0/I)dJ=0.
\end{equation}

In addition to the existing means of determining the location of the phase transition
point, formula (12) permits $J_c$ to be calculated according to the observed velocities
$\Omega_{nc}$.  In Table I we compare the results of processing the experimental data
by different methods for the twenty-eight rotational bands of nonspherical even-even
nuclei that we studied. The $\beta$-vibrational bands of gadolinium and dysprosium are starred.
In column 2 we indicate, of necessity tentatively, the location of the discontinuous
decrease of the angular velocity of rotation.  The value $J_c^{extr}$ is the result
of extrapolation according to formula (3) (see Ref.~1 for more details). The last
column gives the value of the critical spin calculated according to (12).
In calculating the rigid body moment of inertia $I_0$ we used the value of the
nuclear radius obtained in Ref.~1 $r_0 = 1.1\times 10^{-13}$ cm, which is also in
agreement with the data on electron scattering.   It corresponds to the working formula
\begin{equation}\label{15}
    \frac{\hbar^2}{I_0}=\frac{85900}{A^{5/3}} \quad \text{[keV]}.
\end{equation}

\begin{table}
\begin{tabular}{|c|c|c|c|c|c|c|c|}
\multicolumn{8}{c}{\bf Table I}\\
\hline
Nucleus & $J_c^{exp}$ & $J_c^{extr}$  & $J_c^{theor}$ &
Nucleus & $J_c^{exp}$ & $J_c^{extr}$  & $J_c^{theor}$  \\
\hline
$_{56}$Ba$_{68}^{124}$ & $\geq11$ &  & $\leq13.4$ & $_{68}$Er$_{90}^{158}$ & $13-15$ &  & 12.6 \\
$_{56}$Ba$_{70}^{126}$ & $11-13$ &  & $12.5$ &  & $25-27$ & 25.9  & 23.1 \\
$_{58}$Ce$_{70}^{128}$ & $11-13$ &  & $10.6$ & $_{68}$Er$_{92}^{160}$ & $13-15$ &  & 14.7 \\
$_{58}$Ce$_{72}^{130}$ & $9-11$ & 11.0 & $9.7$ & $_{68}$Er$_{94}^{162}$ & $13-15$ & 15.7 & 15.2 \\
$_{58}$Ce$_{74}^{132}$ & $11-13$ &  & $10.3$ & $_{68}$Er$_{96}^{164}$ & $15-17$ &  & 14.4 \\
$_{58}$Ce$_{76}^{134}$ & $9-11$ &  & $9.5$ & $_{70}$Yb$_{94}^{164}$ & $13-15$ & 13.0 & 14.0 \\
$_{64}$Gd$_{90}^{154}$ & $\geq17$ &  & $\leq15.9$ & $_{70}$Yb$_{96}^{166}$ & $13-15$ & 15.5 & 14.5 \\
$_{64}$Gd$_{90}^{154*}$ & $11-13$ & 12.4 & $10.5$ & $_{70}$Yb$_{98}^{168}$ & $15-17$ &  & 18.3 \\
$_{66}$Dy$_{88}^{154}$ & $13-15$ &  & $14.1$ & $_{70}$Yb$_{100}^{170}$ & $15-17$ &  & 18.6 \\
$_{66}$Dy$_{90}^{156}$ & $15-17$ &  & $16.0$ & $_{72}$Hf$_{96}^{168}$ & $13-15$ &  & 13.5 \\
$_{66}$Dy$_{90}^{156*}$ & $11-13$ &  & $9.5$ & $_{72}$Hf$_{98}^{170}$ & $15-17$ &  & 18.6 \\
$_{66}$Dy$_{92}^{158}$ & $15$ & 14.8 & $16.6$ & $_{74}$W$_{96}^{170}$ & $11-13$ & 13.6 & 13.1 \\
$_{66}$Dy$_{94}^{160}$ & $15-17$ &  & $16.0$ & $_{76}$Os$_{106}^{182}$ & $13-15$ & 11.9 & 16.3 \\
$_{68}$Er$_{88}^{156}$ & $11-13$ & 13.2 & $13.8$ & $_{76}$Os$_{108}^{184}$ & $13-15$ & 15.1 & 18.4 \\
 &  &  &  & $_{76}$Os$_{110}^{186}$ & $\geq15$ &  & $\leq17.4$ \\
\hline
\end{tabular}
\end{table}

On the whole the agreement appears to be satisfactory. It should, however, be noted
that the nearness of the magic number $N = 82$ can sometimes manifest itself in very
unexpected ways.   It can be supposed that the second singularity $J^{(2)}$ found
experimentally in the ground state rotational band of Er$^{158}$ (Ref.~4) is due to
this; see also Table I.   In our opinion the construction of more detailed hypotheses
on the nature of the ``intermediate phase" $J_c^{(1)}<J<J_c^{(2)}$ is still somewhat
premature.   Intuition suggests that as the magic nucleus is approached the probability
of similar surprises increases.   In fact, for Er$^{156}$ we already have circumstantial
evidence of this supposition.   According to the data of Ref.~5, after the first phase
transition the moment of inertia, as usual, passes through a minimum, but then rises
very steeply, reaching $I= 1.38 I_0$ for $J = 22$.  The available information indicates
that Er$^{156}$ will also have a second backbending region.

As an illustration of the agreement between the different methods of finding $J_c$ it
is worth noting that the rotational velocity $\hbar\Omega_{nc}$ varies within a fairly
wide range: from 216 keV for W$^{170}$ to 340 keV in the ground state rotational band of
Ba$^{126}$.   The rigid body moment of inertia $I_0$ increases by roughly a factor of
two throughout the entire table.

For nonspherical nuclei sufficiently far from the possible effect of the magic numbers
(see above), experimental data on the upper phase are far from abundant. Therefore,
the possibilities of directly verifying the integral relation (13) or (14) at the
present time are very limited.  Only for W$^{170}$ does the moment of inertia, after
passing through the minimum, approach the rigid body value with an accuracy of about 4\%.
In Fig.~1 we give the graph of the function $1 -I_0/I$ in this case.   The accuracy
with which the integral of this function becomes zero can be considered satisfactory.

\section{Conclusions}

The reason that for tungsten there are few experimental points on a large part of
the moment of inertia curve in the upper phase is the following:  the ratio $j/I_0$ is
not large, amounting only to 1.88.   However, there are nuclei for which the value of
this parameter is much larger.   For Dy$^{158}$, for example, the number of neutrons
$N = 92$ is sufficiently far from the magic number and $j/I_0= 6.76$.   Here further
study of the upper phase makes it possible to construct a more accurate moment of
inertia curve using a considerably larger number of experimental points.   Beginning,
for a rough estimate, from the assumption of similarity to the curve in Fig.~1, we
conclude that in the ground state band of Dy$^{158}$ it is of interest to measure the
location of the rotational levels up to $J\sim 40$, beyond which the moment of inertia
becomes practically the rigid body value.

In addition to the data of Ref.~6 cited in Ref.~1 on the radiation of Ce$^{134}$,
\footnote{Similar data were obtained in [7] for Ce$^{132}$ and Ce$^{130}$.}
verification of the ``area theorem" (13) for different nuclei, including cases with
$j/I_0\gg1$, would give interesting new material for judging the validity of the
theory treating backbending as a macroscopic quantum phenomenon.

We are grateful to I.M.~Pavlichenkov for discussing the results of this study.

\setcounter{equation}{0}

\renewcommand{\theequation}{A.\arabic{equation}}

\section*{Appendix. The condition for rotational stability of a nonspherical nucleus}

The considerations discussed below apply equally to either phase $J\gtrless J_c$.
Let us view the level $J_0$ of the ground state rotational band as minimizing the
total energy of the nucleus for a given value of the conserved angular momentum of
the entire system:
\begin{equation}\label{a1}
    E=\mathrm{min}, \qquad J=J_0.
\end{equation}
Using the Lagrange multiplier method, we shall drop the auxiliary condition and
unconditionally require that
\begin{equation}\label{a2}
    E-\lambda J=\mathrm{min}.
\end{equation}
This can be rewritten as
\begin{equation}\label{a3}
    \delta(E-\lambda J)>0.
\end{equation}
Moving now from the minimum along the actually realized rotational band $E(J)$,
let us calculate the energy change $\delta E$ with accuracy to second-order terms inclusive:
\begin{equation}\label{a4}
    \left\{\left(\frac{dE}{dJ}\right)_{J=J_0}-\lambda\right\}\delta J+\frac12
    \left(\frac{d^2E}{dJ^2}\right)(\delta J)^2>0.
\end{equation}

The value
\begin{equation}\label{a5}
    \lambda=\left(dE/dJ\right)_{J=J_c}
\end{equation}
of the Lagrange multiplier ensures the correct location of the extremum and is a minimum for
\begin{equation}\label{a6}
    \frac{d^2E}{dJ^2}>0.
\end{equation}
Here we have taken into account the fact that the point $J=J_0$
was chosen arbitrarily.   Comparison with formula (1) gives
\begin{equation}\label{a7}
    I>0.
\end{equation}
Therefore, the requirement that the moment of inertia be positive emerges here as
the stability condition.

The question of the rotational stability of a nonspherical nucleus can also be
approached from a somewhat different viewpoint.   Let us as usual denote the
spherical angles giving the orientation of the vector $\mathbf{n}$ (if convenient
it plays the role of the rotational variable, but no separate ``rotational
Hamiltonian" corresponds to it) in the stationary space by $\theta$ and $\phi$.
The state of motion in the angle $\theta$ is, in general, not ``pure."   The
situation with the motion in the azimuthal angle $\phi$ is different: because of
the conservation of the $z$ component of the total angular momentum it corresponds
to the separate wave function
\begin{equation}\label{a8}
    \psi_{rot}=(2\pi)^{-1/2}e^{iM\phi},\quad M=J_z.
\end{equation}
For $J_z=J\gg 1$ the angular momentum is directed along the $z$ axis and the
situation is semiclassical: the fully described state (A.8) of regular precession
can in this limit be viewed as changing into classical motion in a cyclic trajectory,
corresponding to the variation of the azimuthal angle $\phi$.

In order to remove possible doubts about the validity of using a purely mechanical
approach here, we recall that the free rotation of a body in thermodynamic equilibrium
is not accompanied by friction (see Ref. 8, for example).   We shall proceed directly
from the principle of least action
\begin{equation}\label{a9}
    \int_{t_1}^{t_2} Ldt=\mathrm{min}, \quad \phi(t_1)=\phi_1,\quad \phi(t_2)=\phi_2,
\end{equation}
(see Ref.~9, for example).   The extreme values of the angles $\phi_1$ and $\phi_2$
are viewed as constants which are not varied:
$$
    \delta\phi(t_1)=\delta\phi(t_2)=0.
$$
However, let us first consider how the moment of inertia is expressed in terms of
the Lagrangian $L$. We have\footnote{Because of purely classical formulation of
the problem we take $M=\hbar J$ and use from now on usual units.}
$$
    E=E(M),\quad \frac{dE}{dM}=\Omega,\quad L=L(\Omega).
$$
Let us now express the Hamiltonian $E(M)$ in terms of the Lagrangian:
\begin{equation}\label{a10}
E=\Omega dL/d\Omega -L.
\end{equation}
Differentiating (A.10) once with respect to $M$ and cancelling $\Omega$ from both
sides, after some elementary manipulations we easily find
$$
\frac{d^2L}{d\Omega^2}\frac{d^2E}{dM^2}=1.
$$
Comparing this to formula (1), we finally find
\begin{equation}\label{a11}
    I=d^2L/d\Omega^2=(d^2E/dM^2)^{-1}.
\end{equation}
This is precisely the relation between the moment of
inertia and the Lagrangian that we would naturally expect; see also the
preliminary discussion at the beginning of this article.

Let us now return to the principle (A.9), rewriting it in the form
\begin{equation}\label{a12}
    \delta\int_{t_1}^{t_2} L(\Omega)dt>0.
\end{equation}
Free rotation can occur only as uniform rotation, that is, when the angular
velocity is constant:
\begin{equation}\label{a13}
    \Omega_0=\frac{\phi_2-\phi_1}{t_2-t_1}.
\end{equation}
The deviation $\delta\phi(t)$ from this simple law of motion gives rise to a
time-dependent variation of the velocity
$$
\delta\Omega(t)=d\delta\phi(t)/dt.
$$
Let us find the corresponding variation of the action with accuracy to second-order
terms inclusive:
$$
\left(\frac{dL}{d\Omega}\right)_{\Omega=\Omega_0}\int_{t_1}^{t_2} \delta\Omega(t)dt+
\frac12\left(\frac{d^2L}{d\Omega^2}\right)_{\Omega=\Omega_0}
\int_{t_1}^{t_2} [\delta\Omega(t)]^2dt>0.
$$
It is obvious that
$$
\int_{t_1}^{t_2}\delta\Omega(t)dt=\delta\phi(t_2)-\delta\phi(t_1)=0,
\quad \int_{t_1}^{t_2} [\delta\Omega(t)]^2dt>0.
$$
Therefore
\begin{equation}\label{a14}
    \frac{d^2L}{d\Omega^2}>0.
\end{equation}
Comparing this with formula (A.ll), we find the stability condition (A.7).
States with a negative moment of inertia as violating the principle of least action
are unstable and cannot be realized as stationary rotational energy levels of a
nonspherical nucleus.

In conclusion let us clarify the nature of the limiting behavior of the Lagrangian
of the upper phase.  It can be found explicitly in the region near the transition
point, where the moment of inertia has a pole behavior (3).   Substituting this
formula into (A.ll), using formula (30) from Ref.~1 we can express the angular
momentum in terms of the angular velocity and then integrate twice with respect to $\Omega$:
\begin{equation}\label{a15}
    L_n(\Omega)=\frac23(2\hbar j)^{1/2}(\Omega-\Omega_{nc})^{3/2}+I_0\Omega_{nc}\Omega-E_c.
\end{equation}
The constants of integration are here chosen such that the first derivative gives
the angular momentum $M = \hbar J$ and the energy (A.10) coincides for
$\Omega=\Omega_{nc}$ with the nuclear excitation energy $E_c$ at the Curie point.
We see that the Lagrangian of the upper phase has a singularity (a branch point) at
the point of the phase transition $\Omega=\Omega_{nc}$.

\bigskip
\bigskip

\centerline{\bf Figure caption}

\bigskip

Fig.~1.  The integral relation (14) for the W$^{170}$ nucleus.  The area under the
``anomalous" ($I>I_0$) part of the curve is cancelled with an accuracy of about 10\%.

\end{document}